\documentclass[3p, twocolumn]{elsarticle}
\usepackage{siunitx}
\usepackage{booktabs}
\usepackage{graphicx}
\usepackage{subfigure}

\usepackage{amsmath}
\usepackage{pgfplots}
\pgfplotsset{compat=1.16}
\usepackage{xcolor}
\usepackage{hyperref}
\DeclareMathOperator*{\argmax}{arg\,max}
\usepackage[belowskip=-12pt,aboveskip=12pt,font=footnotesize]{caption}
\newcolumntype{L}{>{\centering\arraybackslash}m{8cm}}

\begin{document}

\begin{frontmatter}

\title{Molecular Index Modulation using Convolutional Neural Networks}

\author[1]{Ozgur Kara\corref{cor1}}
\ead{ozgur.kara@boun.edu.tr}
\author[1]{Gokberk Yaylali}
\ead{gokberk.yaylali@boun.edu.tr}
\author[1]{Ali Emre Pusane}
\ead{ali.pusane@boun.edu.tr}
\author[2]{Tuna Tugcu}
\ead{tugcu@boun.edu.tr}

\cortext[cor1]{Corresponding author}

\address[1]{Department of Electrical and Electronics Engineering, Bogazici University, Istanbul 34342, Turkey}
\address[2]{Department of Computer Engineering, NETLAB, Bogazici University, Istanbul 34342, Turkey}


\begin{abstract}
As the potential of molecular communication via diffusion (MCvD) systems at nano-scale communication increases, designing molecular schemes robust to the inevitable effects of molecular interference has become of vital importance. 
There are numerous molecular approaches in literature aiming to mitigate the effects of interference, namely inter-symbol interference. 
Moreover, for molecular multiple-input-multiple-output systems, interference among antennas, namely inter-link interference, becomes of significance. Inspired by the state-of-the-art performances of machine learning algorithms on making decisions, 
we propose a novel approach of a convolutional neural network (CNN)-based architecture. The proposed approach is for a uniquely-designed molecular multiple-input-single-output topology in order to alleviate the damaging effects of molecular interference.  In this study, we compare the performance of the proposed network with that of a  index modulation approach and a symbol-by-symbol maximum likelihood estimation and show that the proposed method yields better performance.

\end{abstract}

\end{frontmatter}

\section{Introduction}

Molecular communication is a novel communication technology that is inspired by the nature. It enables information transmission via micro-machinery by exploiting several biologically-inspired methods. One of the basic approaches of this technology is molecular communication via diffusion (MCvD) \cite{yilmaz2017mcvd}. It is based on the fact that the molecules in a fluid medium flow randomly through space, obeying the laws of Brownian motion. The transmitter (Tx) conveys its message to the receiver (Rx) by encoding the information in a property of the molecular wave of emitted molecules, such as the molecule quantity. Since the molecules propagate according to Brownian motion, they are subject to random propagation. Molecules that are delayed by more than the allotted signaling time are received in subsequent time slots, causing interference among information symbols. This phenomenon is called inter-symbol interference (ISI) and is a significant problem in MCvD systems.

\begin{figure}[h!]
  \centering
  \includegraphics[width=0.7\linewidth]{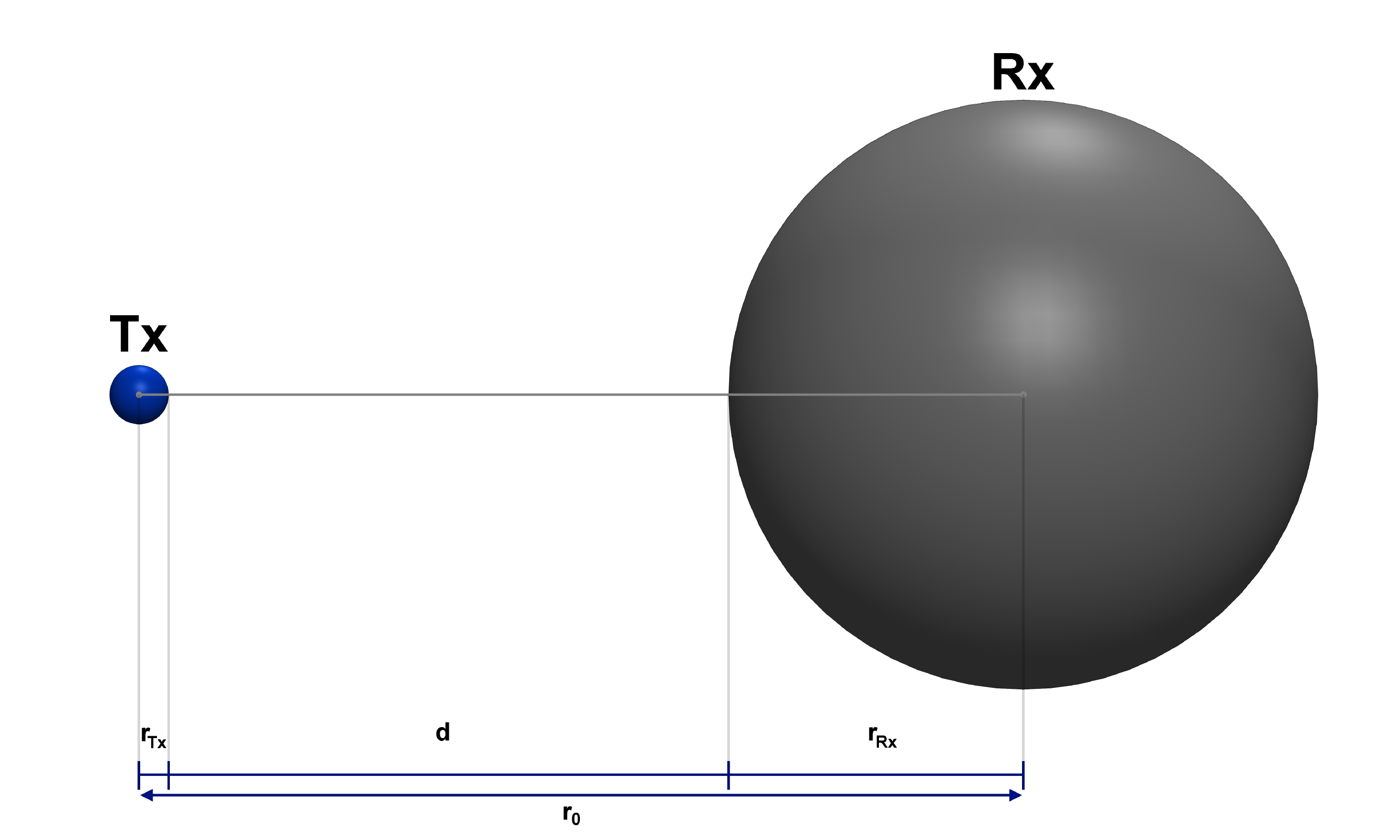}
  \caption{Basic SISO communication scenario for MCvD.}
\end{figure}


Single-input-single-output (SISO) MCvD systems given in Fig. 1 are well-examined molecular communications systems, with their channel characteristics analytically derived in \cite{yilmaz2014three}. The literature presents a multitude of important contributions, including novel modulation methods and advanced reception methods to combat ISI \cite{yilmaz2017mcvd, kabir2015quantity, kuran2011type}. There are several works to mitigate the adverse effects of ISI. In \cite{tepekule2015isimitigation}, ISI mitigation for molecular SISO topologies by adjusting the signal power among symbols to calibrate the residual molecules is conducted. Additionally, this study proposes an energy efficient feedback decision filter to improve communication performance via combating the ISI \cite{tepekule2015isimitigation}. Furthermore, channel coding methods to improve communication quality are proposed in \cite{kislal2020channelcoding}, in which a novel coding method is presented to strengthen the communication by mitigating the residual molecules roaming into the channel. As an example to multiple-type molecular communication schemes, \cite{tepekule2015preequalization} utilizes a modulation scheme, where the residual messenger molecules are reduced by an additional molecule type to mitigate ISI. Despite the many solutions proposed in the literature to reduce the performance-degrading effects of ISI, communication performance --especially at high data rates-- is usually bounded by a severe error floor. Therefore, molecular multiple-input-multiple-output (MIMO) systems are introduced in \cite{yilmaz2016mimo}.

Using multiple Tx/Rx is a conventional method in wireless communications to improve channel quality significantly. However, for ISI-affected applications, such as MCvD, it introduces the risk of further increasing the interference due to inter-link interference (ILI). One of the recently proposed MIMO techniques in MCvD is index modulation (IM) \cite{gursoy2019index, huang2019index}, which is adapted to the molecular communications domain from wireless communications \cite{basar2016index}. In the literature, ISI mitigation techniques are occasionally adapted into molecular MIMO schemes. In \cite{tang2021spatiotemporal}, the index modulation proposed in \cite{gursoy2019index} is enhanced by utilizing a code book and the selection of the antennas depending on the previous symbols. With the proper selection of consecutive antennas, not only ISI, but also ILI is aimed to be mitigated. IM-based communication schemes are good candidates for employing multiple antennas in molecular communication due to their ability to combine the advantages of the spatial domain while keeping the interference at minimum. In index modulation, information is encoded into the selection of the Tx antenna, unlike other MIMO modulation schemes. Rx detects which Tx antenna was used for transmission and decodes the information as its index, accordingly. This scheme enables us to reduce channel use, which results in a cleaner channel with fewer stray molecules and lower interference. 


Employing additional transmitters/receivers always increases implementation complexity. Reducing this complexity while preserving the improved communication performance is one of the main goals in molecular MIMO applications. In order to reduce receiver complexity, a molecular multiple-input-single-output (MISO) topology is proposed in this paper. In contrast to the prevalent MIMO topologies whose receiver regions are separated, MISO topology introduces compact receiver regions on the surface of a single central spherical receiver. Having a single receiver avoids the receiver complexity of molecular MIMO systems.

Receiver regions are located on the central spherical receiver and behave as perfect absorbing receiver surfaces. As done in IM-based molecular MIMO modulations, each equi-areal receiver region corresponds to its conjugate Tx antenna. Suggested receiver topology is able to achieve the single Rx specifications while mimicking multi-receiver Rx designs with its distinctly separated receiver regions. This allows a new approach of molecular MISO topologies in molecular MIMO applications. Possible advantages of using such a single-receiver topology extend from designing systems with multiple single-antenna receivers coexisting in the same channel to creating robust modulation schemes for spatially erroneous topologies. The centralized receiver also allows centralized machine learning-based detection algorithms to outperform conventional detection methods.

In this work, we utilize a machine learning-based single-receiver system for molecular index modulation. The proposed molecular MISO topology is uniquely utilized for molecular IM-based communication schemes. 
Recent studies show that machine learning algorithms are able to learn and benefit from feature representations, especially from multiple time series. In particular, convolutional neural network (CNN)-based architectures have exhibited successful performance on time series classification tasks even though they need to be fed with abundant data during the training phase. Furthermore, unlike artificial neural network (ANN) architectures, convolutional neural networks (CNN) are further able to considerably capture the neighborhood information and cope with the multi-featured data in time series classification problems~\cite{assaf2019mtex, liu2018time}. On the other hand, considering the MISO index modulation scenario mentioned earlier, the receiver has to process multiple time series simultaneously, each of which is correlated with other ones. In the light of these, inspired by the superlative performance of CNN over other typical classification algorithms, we propose the novel use of a CNN-based model to mitigate the error caused by ISI. 

\begin{figure*}
\centering
  \includegraphics[width=0.8\textwidth]{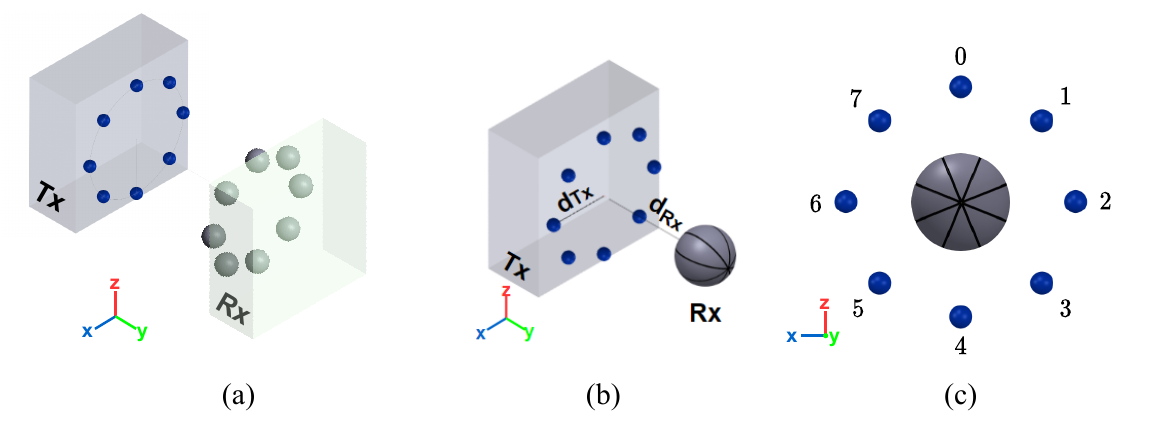}
  \caption{Communication scenarios for $n_{T_x} = 8$ \textbf{(a)} MIMO scenario \textbf{(b)} MISO scenario in 3D view \textbf{(c)} MISO scenario in 2D view. Black lines show the region boundaries on Rx and each region-transmitter conjugate is indexed from 0 to 7, consecutively.}
  \label{fig:miso}
\end{figure*}

The organization of this paper is as follows: Section II demonstrates the proposed system topology. Furthermore, index modulation basics and ML-based detection are briefly explained in this section. In Section III, network details and design specifications of machine learning models are presented. Performances of the methods based on error rates are evaluated in Section IV. Finally, Section V concludes the paper.


\section{System Model}

\subsection{Topology}
The considered MISO topology incorporates $n_{Tx}$ distinct spherical transmitters with radius $r_{Tx}$, whose centers are placed on a uniform circular array (UCA) with the ability to emit molecules into the diffusion channel, as well as a spherical receiver with radius $r_{Rx}$ that is able to absorb the molecules arriving at its surface therewithal record their azimuth and elevation angles. Note that the receiver's center is perfectly placed onto the axis of the UCA at a distance of $d_{Rx}$ from its center.  
The closest distance between the center of the UCA and any transmitter is $d_{Tx}$. This topology is presented in Fig~\ref{fig:miso}.b. Additionally, the receiver sphere is partitioned into $n_{Tx}$ different regions with the purpose that each region becomes a conjugate of the oppositely positioned transmitter. Each Rx region-transmitter conjugate is indexed from 0 to $(n_{T_x}-1)$ consecutively. Region boundaries are further shown in Fig~\ref{fig:miso}.c. The experimental parameters are shown in Table 1. Note that we assume transmitters and receivers are perfectly synchronized.

\subsection{Index Modulation}

\begin{table*}[htbp]
  \centering

  \begin{tabular}{cLc}
    \toprule
    Name & Definition & Value \\
    \midrule
    $n_{Tx}$ & Number of transmitters & 8 \\
    $n_{Rx}$ & Number of receivers & 1 \\
    $r_{Tx}$ & Radius of each transmitter & $\SI{0.5}{\micro\meter}$   \\
    $r_{Rx}$ & Radius of each receiver & $\SI{5}{\micro\meter}$ \\
    $d_{Rx}$ & Distance between the centers' of the receiver and the UCA& $\SI{15.5}{\micro\meter}$ \\
    $d_{Tx}$ & The closest distance between the center of the UCA and each transmitter & $\SI{10}{\micro\meter}$ \\   
    $D$ & Diffusion coefficient & $\SI{79.4}{\micro\meter\squared\per\second}$ \\
    $T$ & Total simulation time & $\SI{5}{\second}$ \\
    \bottomrule
\end{tabular}
\label{tab:sim_param}
  \caption{Simulation Parameters}
\end{table*}

The first adaptations of molecular MIMO modulations into the molecular communication realm aimed increasing overall system throughput at the cost of worse error probabilities \cite{gursoy2019index, huang2019index}. Using the channel multiple times during a symbol duration causes a significant amount of stray molecules --belonging to earlier channel uses-- residing in the channel, which results in heavy interference. To mitigate this effect, it is advantageous to use the channel only once per symbol duration. In this way, information is encoded into the \textit{index} of the intended Tx antenna rather than any other aspect of the molecules, such as quantity \cite{kabir2015quantity}, type \cite{kuran2011type}, or temporal position \cite{garralda2011temporal}. The packet size of the encoded information depends on the number of antennas in use, i.e., for $n_{Tx}$ antennas, information is encoded into $\log_2(n_{Tx})$-bit long packets. Rx antenna collects the messenger molecules through its receiver regions for a defined symbol duration and detects the originating Tx antenna. Information is then extracted through the index of the detected Tx antenna. This general modulation scheme enables reliable information transmission with lower channel use, which directly implies lower interference. In the proposed topology case, receiver regions of Rx are compacted on the surface of the spherical centralized Rx. As aforementioned above, molecules absorbed by Rx are recorded with their azimuth and elevation angles, which enables Rx to perfectly detect the receiver region which a molecule hits during a symbol period. Absorbed molecules are processed to decode the information.

\subsection{Propogation Model}

The proposed machine learning-based approach utilizes the time series of absorbed molecule rates recorded by $n_{Tx}$ regions of the receiver during communication. To train the model, the considered communication scenario is simulated by allowing a randomly selected transmitter to emit $M$ molecules at time $t = (k-1) t_s$, where $k\in\{1,2,...,w\}$, $t_s$ and $w$ denote the symbol period and window number, respectively. For our experiments, $T = \SI{5}\second$ is split into $w\in\{3,4,\dots, 10 \}$ windows per simulation, each of which corresponds to a symbol period of $t_s = \frac{T}{w}$. In other words, 200 simulations are performed for each window $w$ (a total of 1600 simulation runs) to prepare the dataset.  In order to simulate the random movement of the molecules in a driftless fluid environment, total time of communication is divided into time steps of length $\Delta{t}=\SI{d-4}s$. For each symbol duration, i.e., $[(k-1)t_s, kt_s)$, molecules are emitted at the beginning of each window at time $t=(k-1) t_s$.
Note that, in order to make the simulations more realistic, our Tx antennas are designed to be spheres with radius $r_{Tx}$ rather than being point transmitters. Once a transmitter is triggered at time $t=(k-1) t_s$, molecules are generated at the center of the Tx, and start to propagate obeying the rules of diffusion. Moreover, once molecules exit the Tx where they were generated, they are no longer able to diffuse back inside the transmitter; instead, they are reflected.
The position of each molecule in the 3-D space is updated according to
\begin{align} 
& X(t+\Delta{t}) = X(t) + \Delta{X},\\ 
& Y(t+\Delta{t}) = Y(t) + \Delta{Y},\\ 
& Z(t+\Delta{t}) = Z(t) + \Delta{Z},
\end{align}
where $\Delta{X}$, $\Delta{Y}$, and $\Delta{Z}$ are independent and identically distributed Gaussian random variables with $\mu=0$ and $\sigma^2=2D\Delta{t}$, and the diffusion coefficient $D$ is selected as $\SI{79.4}{\micro\meter\squared\per\second}$. Furthermore, if a molecule hits the surface of the receiver, it is absorbed by the receiver and removed from the environment. Note that each receiver region records the number of absorbed molecules at every discrete time of $\Delta{t}=\SI{d-1}{\second}$ and molecule rates are normalized across regions.

Intuitively, the naive approach, namely maximum count decoder (MCD), which predicts the \textit{active} transmitter index for the $k^{th}$ window of the $m^{th}$ sample can be formulated as 
\begin{align}
\hat{y}^m_k = \argmax_i \sum_{j=(k-1)t_s}^{kt_s} x^{m}_{i,j}, 
\end{align}
where $\textbf{X}^m = (x^m_{i,j}),\: i=1,\dots,n_{T_x},\:j=1,\dots,\frac{T}{\Delta{t}} 
$, i.e., $\textbf{X}^{m}$ is the set of the time series for the $m^{th}$ sample containing molecule rates of $n_{T_x}$ regions for each discrete time and is row-wise summed over the interval of the $k^{th}$ window. Thus, the specific region of the receiver that is absorbing the maximum number of molecules can be found, which is identical to the \textit{active} transmitter for the $k^{th}$ window of the $m^{th}$ sample denoted as $\hat{y}^m_k$, i.e., $\hat{\textbf{y}}^m = (\hat{y}^m_k),\: k=1,\dots,w $.





\begin{figure*}
  \centering
  \includegraphics[width=1\textwidth]{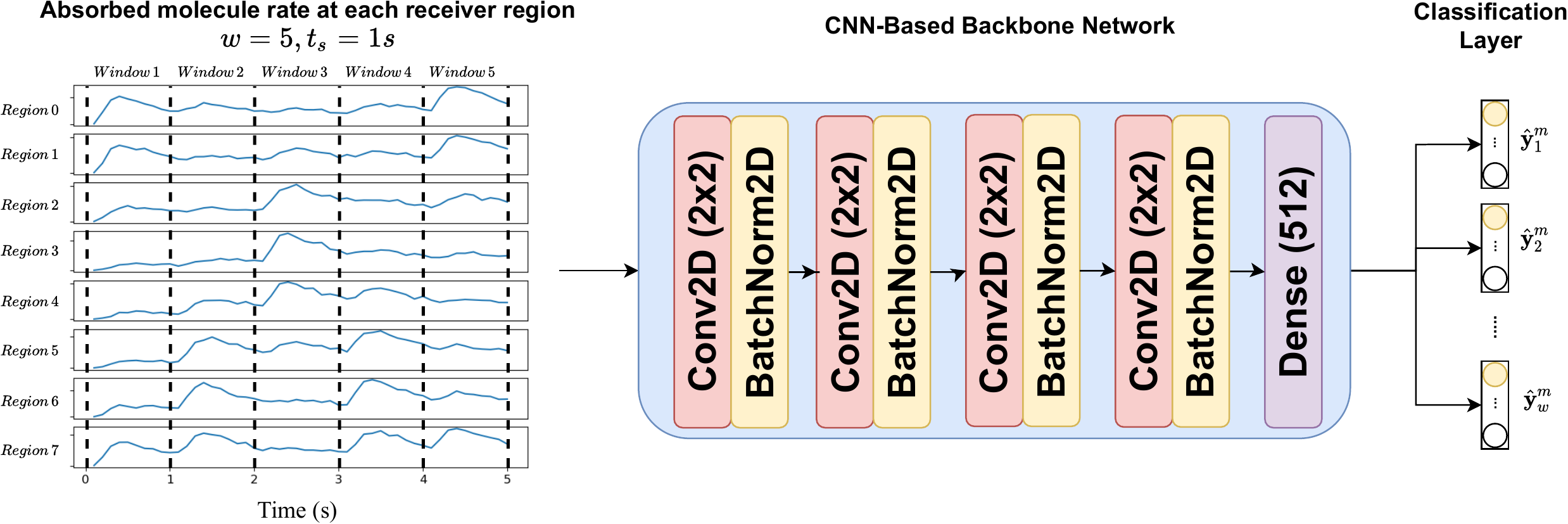}
  \caption{Outline of the proposed model for $m^{th}$ sample where $w=5$ and $t_s = \SI{5}{\second}$. At each window following indexed transmitters emitted molecules: $0-6-3-6-0$.}
  \label{fig:cnn_model}
\end{figure*}

\section{Machine Learning Model}

The task of detecting symbols at the receiver side turns into a multivariate time series classification problem. As each receiver region records the normalized rate of molecule distribution, the network is fed with $n_{T_x} = 8$ different time series simultaneously. 

Each transmitter is indexed with a value from 0 to 7 consecutively (see Fig.~\ref{fig:miso}-c). In order to predict the correct transmitter index for each window, a CNN-based neural network architecture is designed encompassing two main parts. The first is the CNN-based model (given in Fig.~\ref{fig:cnn_model}), which is the backbone of the network, and the second is a classifier layer that is specialized for each distinct window number $w\in\{3,4,\dots, 10 \}$.  Based on the selection among eight different windows, eight different models, which are differentiated according to their number of \textit{sub-heads} at the classifier layer, are implemented. The classifier layer of each model is separated into $w$ \textit{sub-heads}, each of which comprises eight neurons being the probabilities of having used the corresponding transmitter. 
In other words, the algorithm can be formulated as $\mathcal{F}_w(\textbf{X}^m\mid\textbf{y}^m)$ where $\textbf{X}^m$ is a set of time series with dimensions $(n_{T_x},\frac{T}{\Delta{t}})$, $\textbf{y}^m$ is the one-hot encoded ground truth label with dimensions $(w,n_{T_x})$, $w$ is the window number, and superscript $m$ denotes the $m^{th}$ sample.
The objective of the machine learning model $\mathcal{F}_w$ is to learn the relation between the molecule distribution over eight regions of the receiver as well as the \textit{active} transmitter for each window.
Note that, for each window number $w$, the model is trained using the data generated with $M=10^5$ molecules. Each model is evaluated on its ability to correctly decode the transmissions done with varying molecule amounts $M\in\{750,1000,\dots, 3250 \}$. 

The backbone network is composed of four convolutional layers with kernel sizes $(2,2)$ having $512$ filters connected to a $512$-dimensional dense layer after the convolutional layers. As an activation function, rectified linear unit (ReLU) is attached after each convolution block as well as after each dense layer. In order to alleviate the effects of overfitting, the batch normalization layer is utilized after each convolutional layer, which is beneficial in making neural networks faster and more stable by re-centering and scaling the mini-batches based on their mean and standard deviation values~\cite{ioffe2015batch}. Softmax is applied after each \textit{sub-head} in order to convert the linear outputs to a probability distribution. Particularly, the $i^{th}$ element of the vector $\hat{\textbf{y}}_n^m$, $\hat{y}_{n,i}^m$, denotes the probability of the $i^{th}$ transmitter that is \textit{active} for the $n^{th}$ time window/symbol and the $m^{th}$ sample.

To train such a \textit{multi-head} architecture for a classification task, categorical cross entropy loss function is applied through each of the \textit{sub-heads}. To calculate the total model loss $\mathcal{L}_{total}$ per mini-batch, all losses found for individual \textit{sub-heads}, given as

\begin{align}
\mathcal{L}_n(\hat{\textbf{y}}_n,\textbf{y}_n) = -\frac{1}{B}\sum_{m=1}^B\sum_{i=1}^{n_{Tx}}y_{n,i}^m \log { \left( \hat{y}^m_{n,i} \right)},  
\label{eq:loss_for_subhead}
\end{align}
are averaged over the windows as
 \begin{align}
\mathcal{L}_{total} = \frac{1}{w}\sum_{n=1}^{w}\mathcal{L}_n(\hat{\textbf{y}}_n,\textbf{y}_n).
\label{eq:average_loss}
\end{align}
Here, $\mathcal{L}_n(\hat{\textbf{y}}_n, \textbf{y}_n)$ is the calculated loss for the $n^{th}$ \textit{sub-head}, $B$ is the mini-batch size, $n_{Tx}$ denotes the number of transmitters (which always equals to $8$ in our case), and $\hat{y}^m_{n,i}$ and $y_{n,i}^m$ denote the predicted value and one-hot encoded ground-truth label of the $i^{th}$ neuron at the $n^{th}$ window/\textit{sub-head} for the $m^{th}$ sample in the mini batch, respectively. 
 
All models are trained with the stochastic gradient descent algorithm~\cite{robbins1951stochastic}, using Adam~\cite{kingma2014adam} as an optimizer with a learning rate of $0.001$, and the network is trained for $200$ epochs with a batch size of $64$. Hyper-parameter values are selected to be the best-performing ones based on hyper-parameter search experiments. In addition, using complicated models with more layers shows no significant improvement in the performance, albeit their extra complexity.

Practically speaking, machine learning methods are somewhat similar to other decoder/demodulation techniques since they can be thought of as a different representation of multiplication and summation operations. The key point is that the machine learning model should be trained outside of the nanomachine prior to physical implementation. The dataset can be produced either by simulations or real-world experiments, which is in our case obtained by simulations. Then, this pretrained model can be embedded into the receiver, which will perform the classification using multiple summation and multiplication operations.

\section{Performance Results}
The communication performance of the suggested MISO topology in Section II-A is evaluated through Monte Carlo simulations. For performance evaluations, the maximum count decoder, formulated in (4), is employed as the baseline approach for comparisons. To further evaluate the performance of the proposed method, our method is compared to the symbol-by-symbol maximum likelihood estimator (MLE) method that is explained in \MakeUppercase{\romannumeral 7}-C section of \cite{gursoy2019index}. The receiver that is designed with this method is capable of storing the previous decisions. Then, it calculates the estimated past arrival mean and variances as 
	\begin{align*}
	    \hat\mu_{j,past}[k] = \sum_{z=0}^{k-1}\sum_{i=0}^{n_{Tx}-1} \hat{s}_i[z]h_{i,j}[k-z+1]
	\end{align*}
	and
	\begin{align*}
	    &\hat\sigma^2_{j,past}[k] = \\ &\sum_{z=0}^{k-1}\sum_{i=0}^{n_{Tx}-1} \hat{s}_i[z]h_{i,j}[k-z+1](1-h_{i,j}[k-z+1])
	\end{align*}
using the previous decisions and channel coefficients ($h_{i,j}[k])$, which are easily obtained by running a simulation with a large molecule number to make it more precise. Note that $x[k]$ denotes the activated antenna for $k^{th}$ time instant, $n_{Tx}$ denotes the number of transmitters. Then, $\hat{s}_i[z]=\frac{\log n_{Tx}}{2}M^{Tx}$ if $\hat{x}[k] = i$, and is zero otherwise. After determining the past arrival values, the decoder calculates the estimated mean and variance vectors (for all Tx antennas) as
	\begin{align*}
	    (\boldsymbol{\hat\mu_i[k]})_j  = \hat\mu_{j,past}[k] + s_{MSSK}h_{i,j}[1]
 	\end{align*}
 	and similarly
 		\begin{align*}
	    (\boldsymbol{\hat\sigma^2_{i}[k]})_j = \hat\sigma^2_{j,past}[k] + s_{MSSK}h_{i,j}[1](1-h_{i,j}[1]) 
 	\end{align*}
for the given $i^{th}$ active Tx antenna, where $s_{MSSK}=\frac{\log n_{Tx}}{2}M^{Tx}$. After all, the objective is to find the index $i$ which maximizes the log-likelihood function applied to our receiver given the received number of molecules for that window.
Then, the index becomes
 \begin{align*}
     \hat{i} = \argmax_{i}\sum_{j=0}^{n_{Rx}-1}\ln & \left(\frac{1}{\sqrt{2\pi (\boldsymbol{\hat\sigma^2_{i}[k]})_j}}\right)\\ &-\frac{(R_j[k]-(\boldsymbol{\hat\mu_i[k]})_j)^2}{2(\boldsymbol{\hat\sigma^2_{i}[k]})_j},
 \end{align*}
 where $R_j[k]$ is the number of received molecules and $n_{Rx}$ denotes the number of regions on the Rx antenna.

The unique contribution of this paper is the proposed machine learning model, and the required evaluations are conducted with the help of computer simulations. Communication simulations are conducted for both high and low data rates. Also, simulation results for varying bit durations $t_b$ are given in order to show the convergence of performance under the ISI and ILI effects.

\begin{figure}[h!]
\begin{center}
  \includegraphics[width=1\linewidth]{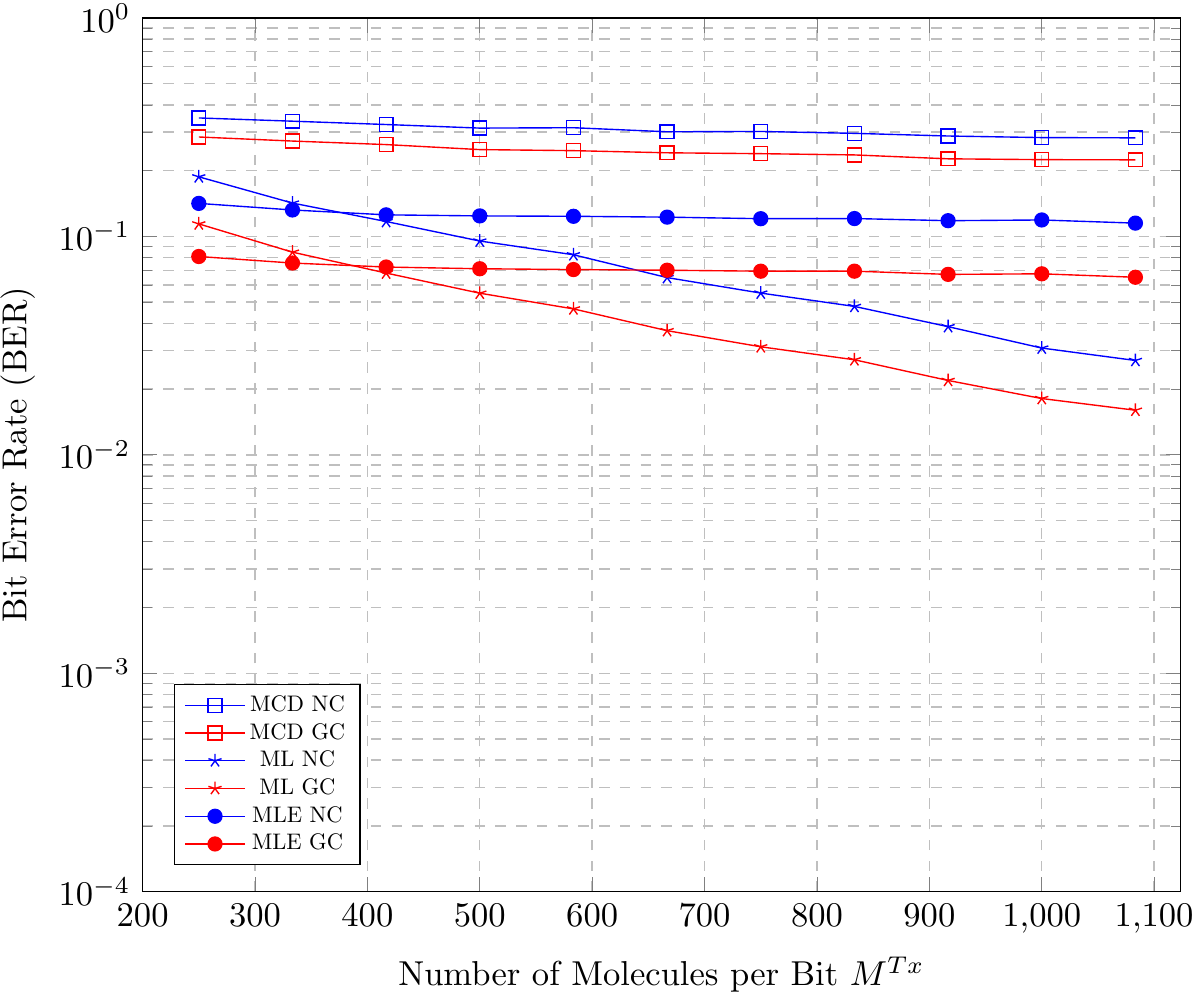}
    \caption{Natural Coding (NC) and Gray Coding (GC) bit error rate for Maximum Count Decoder (MCD), Machine Learning (ML), Symbol-by-symbol Maximum Likelihood Estimation (MLE) approaches with $t_b = \SI{0.166}{\second}$. }
    \label{fig:window_0166}
\end{center}
\end{figure}

In Fig.\ref{fig:window_0166}, simulation results for high data rate scenario are presented. Since the receiver regions are positioned adjacently as segments of a spherical receiver, the probability of misreceived molecules (and the amount of ILI) is significantly high. This causes a significant number of molecules to be absorbed by adjacent receiver regions,  resulting in ILI. For a high data rate communication, ISI becomes the dominant interference problem for the molecular realm along with ILI. MCD is prone to error due to significant interference. Ideally, the maximum-likelihood sequence estimator would perform optimally in term of error rates. However, it is not feasible to implement such an estimator for a nano-scaled communication system due to its high complexity. The symbol-by-symbol maximum-likelihood estimator shows a better performance compared to MCD. The proposed method shows strong performance compared to the symbol-by-symbol MLE at high data rate scenario due to its ability to learn interference patterns with higher success than the symbol-by-symbol MLE.

\begin{figure}[h!]
\begin{center}
  \includegraphics[width=1\linewidth]{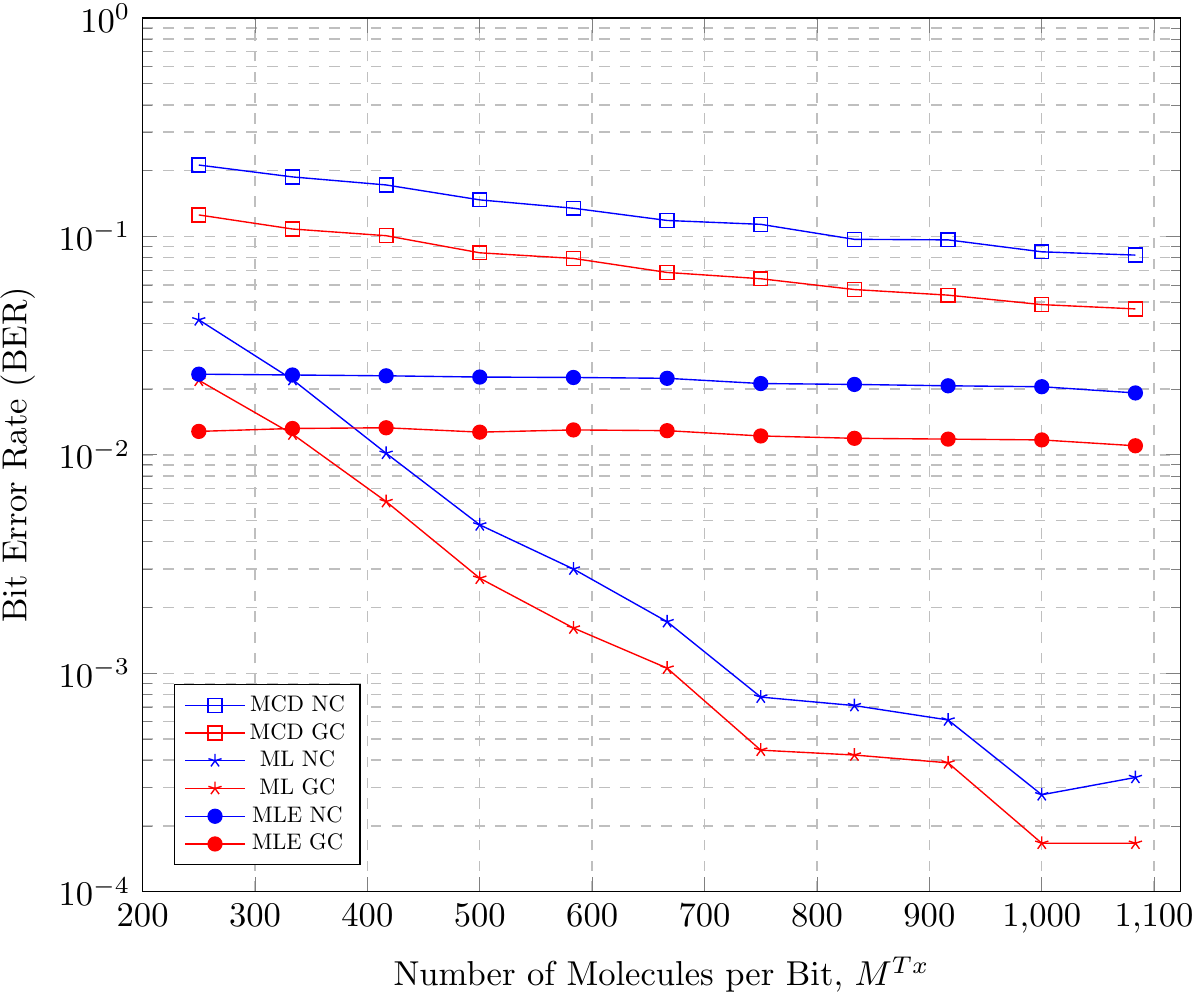}
    \caption{Natural Coding (NC) and Gray Coding (GC) bit error rate for Maximum Count Decoder (MCD), Machine Learning (ML), Symbol-by-symbol Maximum Likelihood Estimation (MLE) approaches with $t_b = \SI{0.555}{\second}$.}
    \label{fig:window_0555}
    \end{center}
\end{figure}

For low data rate scenario given in Fig. 5, ILI becomes the dominant interference source. MCD is prone to errors due to considerable ILI. However, the proposed machine learning method significantly dominates both MCD and the maximum-likelihood estimation in bit error rate performance. The ability to learn and recognize interference patterns of the proposed method is enhanced when low data rates result in lower ISI.

\begin{figure}[h!]
\begin{center}
  \includegraphics[width=1\linewidth]{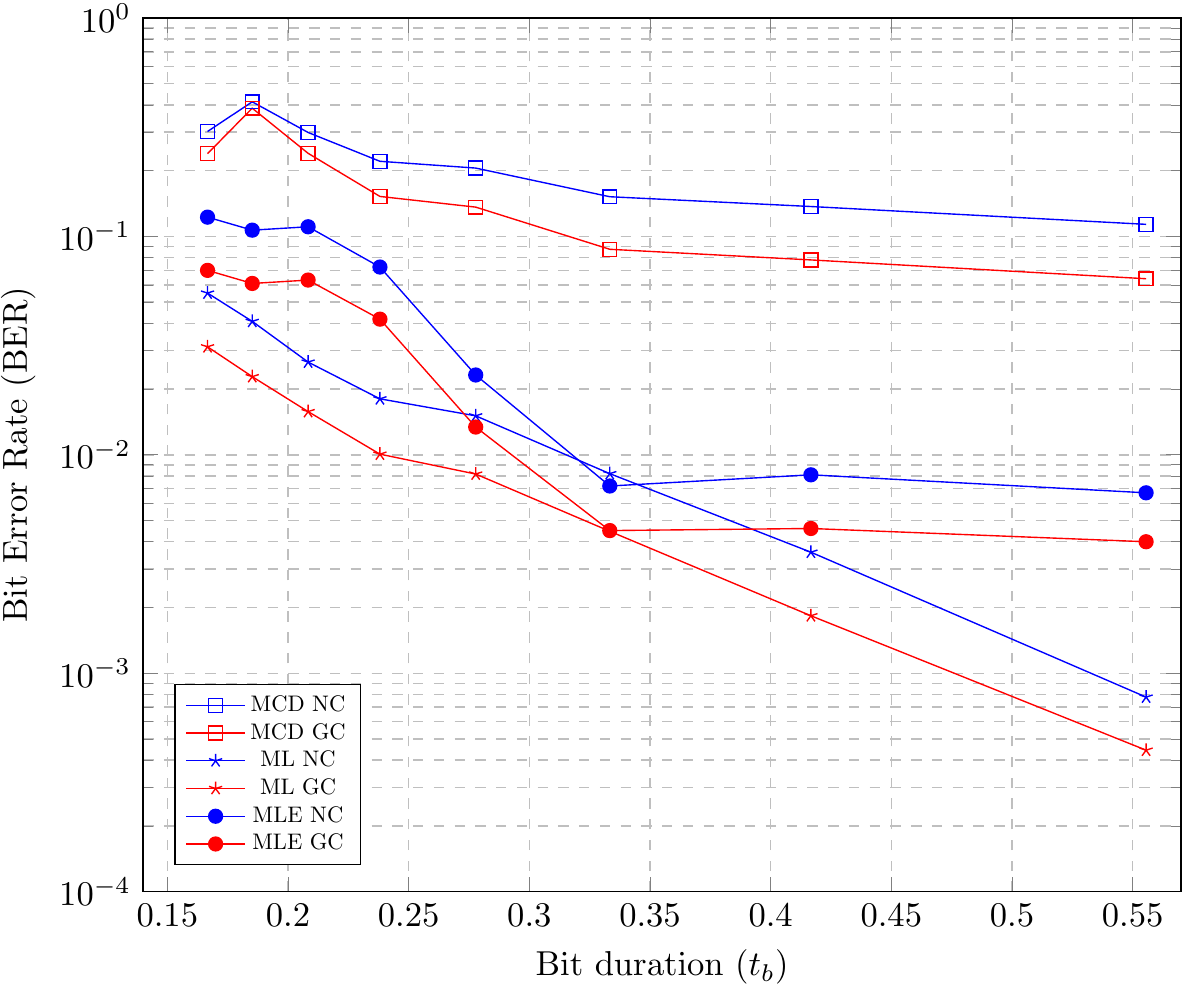}
    \caption{Natural Coding (NC) and Gray Coding (GC) bit error rate for Maximum Count Decoder (MCD), Machine Learning (ML), Symbol-by-symbol Maximum Likelihood Estimation (MLE) approaches with varying $t_b$ and $750$ molecules per bit.}
    \label{fig:tb_plot}
    \end{center}
\end{figure}

As discussed before, the molecular MISO topology is particularly vulnerable to ILI, as a result, the performance improvement is limited when $t_b$ increases. In the varying $t_b$ scenario given in Fig. 6, where the number of molecules released per bit, $M^{Tx}$, is fixed, the topology shows its natural limits as the error performance improvement halts due to ILI, error performance improvement halts, resulting in an error floor. The MCD has a significantly high error floor. On the other hand, the proposed machine learning method continues to adapt under increasing $t_b$ and shows no error-flooring. This strength comes from the natural ability of the proposed approach, which indicates continuous adaptation to learn patterns and recognize interference sequences. Nearly at all $t_b$ scenarios, the proposed method outperforms both the MCD and the symbol-by-symbol maximum-likelihood estimator.

\section{Discussion on Different Scenarios}
In this section, we discuss other practical scenarios (e.g. using a higher number of transmitter antennas, varying channel conditions, different receiver sizes, etc.) and their expected results based on our findings.

As the number of transmitters increases, the surface area and the angle of each region on the receiver that corresponds to the transmitters will decrease. As a result, an increase in the inter link interference (ILI) is expected since molecules are more likely to hit and be absorbed by neighboring regions. To manage this, a larger receiver should be used to increase the surface region, which increases the implementation costs. Furthermore, if more transmitters are used in our scenario, the radius of uniform circular array (UCA) ($d_{Rx}$) should be increased to fit all of these transmitters in a circular shape, since transmitters are assumed to be little spheres rather than being points. Hence, transmitters and receiver move away from each other, which in turn increases the required time that molecules reach to receiver. Thus, communication slows down considerably and the symbol duration should be extended to compensate these changes to keep ISI from increasing.

The same logic applies when channel conditions, i.e., diffusion coefficient, changes. Since the square root of the diffusion coefficient is proportional to the variance of the displacement of molecules, it affects the speed of communication, which is directly related to ISI. Meanwhile, since the variance increases, the molecules are liable to hit neighboring regions. Hence, ILI is likely to cause further errors.

Again, using a larger receiver allows a greater surface region for each transmitter and the surface of the receiver and each transmitter gets closer to each other. Hence, both ILI and ISI errors will be alleviated.


\section{Conclusion and Future Work}
In this study, a machine learning-based molecular index modulation scheme for a newly-proposed molecular MISO topology has been introduced. Said topology is new to the molecular communication realm and enables reducing receiver complexity drastically. It is uniquely implemented on molecular IM-based communication schemes. Since the receiver regions are compacted on the single central receiver, the implementation complexity of the proposed scheme is realizable. It has been shown that the proposed molecular MISO topology is competent to provide adequate communication performance. Moreover, as aforementioned before, molecular MISO topology is able to fulfill the index modulation potential suggested by molecular MIMO topologies, which allows conducting IM-based modulation schemes using molecular MISO topology without losing the improved communication performance superiority promised by index modulation. Presented results are able to show that said topology is performing satisfactory under basic IM-based modulation schemes, namely the MCD, for this paper.

Another unique contribution of this paper is the proposed machine learning-based molecular index modulation scheme. The adaptive nature of machine learning methods enables overcoming communicative obstacles of molecular interference, namely ISI and ILI. It has been shown that machine learning-based modulation methods are prone to recognize interference patterns that are caused by the random nature of the molecular communication realm. The experimental results support the fact that the proposed machine learning-based modulation outperforms the MCD of the basic index modulation scheme conducted on molecular MISO topology at both low and high data rates. Due to our concerns for a fair comparison, a symbol-by-symbol MLE is performed. Based on the previous decisions and the channel characteristics knowledge, MLE detects the most likely antenna as the originating antenna for the considered symbol duration. The proposed machine learning scheme outperforms the MLE at almost all symbol power levels and has a much more promising betterment curve concerning the symbol power. Due to the nature of topology, molecular MISO topology is prone to suffer heavily from ILI. The proposed machine learning method suggests an effective modulation scheme with significant ISI mitigation, without depending on the perfect channel information.

The main goal is to introduce machine learning-based molecular index modulation schemes for newly-proposed MISO topologies, possible spatial misalignments of antennas, angular deviations or other imperfection assumptions are outside of the scope of this paper. Possible extensions of this study such as designing robust nano-networking hub with multiple receiver users are left as future works. Moreover, aforesaid spatial imperfections regarded for molecular MISO topology can be overcome, since rotational shifts can be considered straight-forward tasks for machine learning schemes. Therefore, with the help of such machine learning schemes, such further problems are effortless to provide a solution. Additionally, the future work entails the development of different machine learning models for different scenarios.

\section{Acknowledgements}
This work was supported in part by the Scientific and Technical Research Council of Turkey (TUBITAK) under Grant 119E190.

\bibliographystyle{elsarticle-num} 
\bibliography{main.bib}

\end{document}